\documentclass[prb,twocolumn,showpacs]{revtex4-1}

\usepackage{amsmath,amssymb,graphicx}

\usepackage{color}

\newcommand{\ii}{\mathrm{i}}

\begin{document}

\title{A variational discrete variable representation for excitons on a lattice}

\author{A. Alvermann}
\author{P. B. Littlewood}
\affiliation{Theory of Condensed Matter, Cavendish Laboratory, Cambridge CB3 0HE, United Kingdom}
\author{H. Fehske}
\affiliation{Institut f\"ur Physik, Ernst-Moritz-Arndt-Universit\"at, 17487 Greifswald, Germany}

\begin{abstract}
We construct numerical basis function sets on a lattice,
whose spatial extension is scalable from single lattice sites to the continuum limit.

They allow us to compute small and large bound states with comparable, moderate effort. 
Adopting concepts of discrete variable representations, a diagonal form of the potential term is achieved through a unitary transformation to Gaussian quadrature points.
Thereby the computational effort in three dimensions scales as the fourth instead of the sixth power of the number of basis functions along each axis,
such that it is reduced by two orders of magnitude in realistic examples.
As an improvement over standard discrete variable representations, our construction preserves  the variational principle. It allows for the calculation of binding energies, wave functions, and excitation spectra.
We use this technique to study central-cell corrections for excitons beyond the continuum approximation.
A discussion of the mass and spectrum of the yellow exciton series in the cuprous oxide, which does not follow the hydrogenic Rydberg series of Mott-Wannier excitons, is given on the basis of a simple lattice model.
\end{abstract}


\maketitle

\section{Introduction}

The properties of elementary excitations in solids can often be understood in a continuum approximation.
A prominent example are bound electron-hole pairs in semiconductors, the excitons~\cite{Kno63}.
Mott-Wannier~\cite{Wa37,Mo38} excitons with large radius, found in materials such as silicon or gallium arsenide, are in first but rather good approximation described in terms of hydrogen atoms.

For smaller exciton radius, comparable with the lattice constant, deviations from the hydrogen-like properties occur.
Central-cell corrections become important, which account for the possibility of finding 
electron and hole in the same unit cell and include non-parabolic dispersions
and modifications of the $1/r$-Coulomb potential~\cite{KCB97}.
A prototypical material is the cuprous oxide $\mathrm{Cu}_2 \mathrm{O}$,
which receives constant attention in the search for an excitonic Bose-Einstein condensate~\cite{MS00,SS10}.
It features an interesting property:
The mass of the ``yellow'' 1s excitons ($2.6 m_0$, where $m_0$ is the free electron mass) is larger than the sum of the electron ($1.0 m_0$) and hole mass ($0.7 m_0$)~\cite{BFSBSN07}.
The mass enhancement, after all by $50 \%$, indicates the breakdown of the continuum approximation.

It is our intention to study central-cell corrections for excitons starting from a microscopic lattice model. In the present paper we restrict ourselves to a simple two-band model.
Despite its limitations the discussion will provide us with a first understanding and set the reference for future work.
Improved studies, e.g. with the inclusion of realistic ab-initio band structures, and extensions to biexcitons and exciton-exciton scattering are under current investigation.

The study of excitons provides motivation for the development of numerical methods for few-particle systems on lattices.
For small exciton radius,
eigenstates can be obtained from a ``plain'' lattice calculation,
where the numerical wave function is restricted to a finite number of lattice sites.
With maximal distance $L$ between electron and hole,
wave function values at $(2L+1)^3$ lattice sites must be stored in memory for a three-dimensional (3D) problem.
As a consequence of the $L^3$ scaling the resource consumption of such calculations increases rapidly with the exciton radius. 
Already for a radius of, say, 10 lattice sites we must deal with about $10^6$ wave function values.
Excited states are obtained with even higher effort. 
For biexcitons with four particles, the scaling is $\propto L^9$.

The above numbers indicate that a different approach is needed.
In the present work we introduce a variational lattice formulation of discrete variable representations (DVR) developed for molecular physics and theoretical chemistry applications~\cite{HEG65,DC68,LHL85,BH86}.
The wave function is represented in a product basis of sine functions, whose spatial width is a free parameter that is varied in order to minimize the energy and thus optimize the numerical wave function.
For small basis function width, this approach reduces to a plain lattice calculation.
Allowing the basis function width to grow the transition to the continuum limit
is addressed with constant effort which is independent of the wave function radius.

For a decisive reduction of the computational effort we rely on the DVR idea of using Gaussian quadrature for the potential term in the Hamilton operator.
With $N$ basis functions along each coordinate axis in 3D,
a straightforward variational calculation in the sine basis requires all 
$N_\mathrm{tot}^2$ potential matrix elements between the $N_\mathrm{tot}= N^3$ basis functions. These matrix elements have to be calculated and then used in each application of the Hamilton operator.
With the DVR, the potential term is represented by a diagonal matrix and thus requires only $N_\mathrm{tot}$ matrix elements.
Since the kinetic energy is separable in appropriate coordinates, the total effort scales only as $N^4=N_\mathrm{tot}^{4/3}$ instead of $N^6=N_\mathrm{tot}^2$.
For $N=100$, the effort is reduced by a factor of the order $10000$.

We deviate in two points from the standard DVR formulation.
First, we use a lattice function basis.
Second, we adapt the DVR construction to circumvent its single drawback,
the violation of the variational principle through the Gaussian approximation of the potential.
In our formulation, the DVR strictly obeys the variational principle.
This is important because it allows for the determination of  the optimal basis function width through energy minimization.
Since we do not know the wave function size, e.g. the excitonic lattice Bohr radius, in advance,
the selection of a suitable basis function width would otherwise be difficult and error-prone.
With these two modifications, our approach covers the entire range of small bound states occupying few lattice sites through the intermediate regime up to the limit  of large weakly bound states in the continuum.

The paper is organized as follows.
In Sec.~\ref{sec:Model} we introduce the exciton lattice model for the present investigation,
and derive the central relation between exciton mass and kinetic energy.
In Sec.~\ref{sec:Method} the variational sine basis is defined, the variational DVR is explained and applied to two test examples.
In Sec.~\ref{sec:Exciton} we study central-cell corrections for small radius excitons using results from variational DVR calculations for the lattice model.
After a discussion of the yellow exciton series in the cuprous oxide, we summarize our findings in Sec.~\ref{sec:Conc}.

\section{The exciton model}\label{sec:Model}

We study 3D excitons within a simple two-band model on a cubic lattice,
with a cosine dispersion 
\begin{equation}\label{Disp}
\begin{split}
E_{e/h}(\mathbf{k}) &= \frac{\hbar^2}{a^2 m_{e/h}} \sum_{i=x,y,z} (1- \cos a k_i) \\
 &= \frac{\hbar^2 k^2}{2 m_{e/h}} -  \frac{\hbar^2 a^2 (k_x^4+k_y^4+k_z^4)}{24 m_{e/h}} + \dots
\end{split}
\end{equation}
for the conduction ($E_e$) and valence ($E_h$) band.
Instead of the electron mass $m_e$ and hole mass $m_h$,
we will also use the reduced ($m_r$) and total ($M$) electron-hole mass
\begin{equation}
m_r^{-1} = m_e^{-1} + m_h^{-1} \;,  
M=m_e + m_h \;.
\end{equation}
The parameter $a$, with physical dimension `length', determines the typical exciton radius below which central-cell corrections are significant.
It could be identified as the lattice constant for realistic band dispersion,
but should be considered as an effective model parameter in the present simple treatment, similar to Ref.~\onlinecite{KCB97}.

\subsection{The lattice model}

The lattice Hamilton operator is given as the sum
\begin{equation}
 H = T_e + T_h + U 
\end{equation}
of the kinetic energy of the electron
\begin{equation}\label{Te}
 T_e = \frac{\hbar^2}{2 a^2 m_e} \Big( 6 - \sum_{\substack{i=x,y,z \\ \delta=\pm 1}} \sum_{\mathbf{r}} c^\dagger_{\mathbf{r}+ \delta \mathbf{e}_i}   c^{}_\mathbf{r} \Big) \;,
\end{equation}
the kinetic energy of the hole
\begin{equation}\label{Th}
 T_h = \frac{\hbar^2}{2 a^2 m_h} \Big(6 -  \sum_{\substack{i=x,y,z \\ \delta=\pm 1 }}  \sum_{\mathbf{r}} h^\dagger_{\mathbf{r}+\delta \mathbf{e}_i}   h^{}_\mathbf{r}  \Big) \;,
\end{equation}
and the attractive Coulomb interaction 
\begin{equation}
 U = \sum_{\mathbf{r} \mathbf{r}'}  U(\mathbf{r}-\mathbf{r'})  
 c^\dagger_\mathbf{r}   c^{}_\mathbf{r}
 h^\dagger_\mathbf{r'}   h^{}_\mathbf{r'}
\end{equation}
between both, with
\begin{equation}\label{UCoul}
 U(\mathbf{r}) = \begin{cases} -\dfrac{e^2}{\epsilon a |\mathbf{r}|} \quad & \text{ if } \mathbf{r} \ne 0 \;, \\[1em]
  -V \dfrac{e^2}{\epsilon a} & \text{ if } \mathbf{r} = 0 \;.
  \end{cases} 
\end{equation}
In these expressions,
$c^{(\dagger)}_\mathbf{r}$ and $h^{(\dagger)}_\mathbf{r}$ denote fermionic operators for
an electron or hole at lattice site $\mathbf{r}$.
The lattice sites are indexed by integer numbers, i.e. $r_i \in \mathbb{Z}$
(this explains the appearance of the parameter $a$ in the Coulomb interaction).
The unit vector along each axis is denoted by $\mathbf{e}_i$.
The Hamilton operator in the form given has five parameters, some of which are redundant as will be seen later:
The electron/hole masses $m_{e/h}$, the (effective) lattice constant $a$,
the dielectric constant $\epsilon$, and the local Coulomb factor $V$.
While $\epsilon$ characterizes the long-range part of Coulomb interaction, the parameter $V$ describes the relative strength of electron and hole interaction in the same unit cell.
It should be $V > 1$ since $|U(0)| > |U(\mathbf{r}Ê\ne 0)|$.

\subsection{Separation of center-of-mass motion}

The exciton wave function can be written as 
\begin{equation}
  |\psi\rangle = \sum_{\mathbf{r} \mathbf{r}'} \psi(\mathbf{r},\mathbf{r}') c^\dagger_\mathbf{r} h^\dagger_{\mathbf{r}'} |\mathrm{vac}\rangle
\end{equation}
where $|\mathrm{vac}\rangle$ denotes the semiconductor ground state with filled valence and empty conduction band.
We are interested in wave functions with definite exciton momentum $\hbar \mathbf{K}$.
Translational invariance requires
\begin{equation}
\psi(\mathbf{r}+\mathbf{R},\mathbf{r}'+\mathbf{R}) = e^{\ii a \mathbf{K} \cdot \mathbf{R}} \psi(\mathbf{r},\mathbf{r}') \;,
\end{equation}
which allows for separation of  the center-of-mass motion through the ansatz
\begin{equation}
 \psi(\mathbf{r},\mathbf{r}') = e^{\ii a \mathbf{k}_e \cdot \mathbf{r} + \ii a \mathbf{k}_h \cdot \mathbf{r}'} \phi(\mathbf{r}-\mathbf{r}') \;. 
\end{equation}
Every combination $\mathbf{k}_e + \mathbf{k}_h = \mathbf{K}$ is allowed,
and any two choices are related by a unitary transformation.

The wave function of relative electron-hole motion $\phi(\mathbf{r})$
obeys the effective one-particle Schr\"odinger equation
\begin{equation}
\begin{split}
 E \phi(\mathbf{r}) &= \frac{3 \hbar^2}{a^2 m_r } \phi(\mathbf{r}) \\
 &-  \frac{\hbar^2}{2 a^2} \sum_{\substack{i=x,y,z \\ \delta=\pm 1}} \Big(  \frac{e^{\ii a \delta \mathbf{k}_e \cdot \mathbf{e}_i}}{m_e}   
 + \frac{e^{-\ii a \delta \mathbf{k}_h \cdot \mathbf{e}_i}}{m_h}  \Big)  \phi(\mathbf{r}+\delta \mathbf{e}_i)
 \\ & + U(\mathbf{r}) \phi(\mathbf{r}) \;.
 \end{split}
\end{equation}
In general, this equation is complex.
For the simple cosine dispersion assumed here, however, the choice 
\begin{equation}
 \frac{m_e}{m_h}=\frac{\sin a \mathbf{k}_e \cdot \mathbf{e}_i }{\sin a \mathbf{k}_h \cdot \mathbf{e}_i } \qquad (\text{for } i=x,y,z) \;,
\end{equation}
that is 
\begin{equation}
\begin{split}
a \mathbf{k}_e \cdot \mathbf{e}_i &= \arctan \frac{ \sin a k_i}{\cos a k_i + \frac{m_h}{m_e}}
 \;, \\
 a \mathbf{k}_h  \cdot \mathbf{e}_i &= \arctan \frac{ \sin a k_i}{\cos a k_i + \frac{m_e}{m_h}} \;,
 \end{split} 
\end{equation}
leads to a Schr\"odinger equation $E \phi = \tilde{H} \phi$ 
with the real Hamilton operator
\begin{equation}\label{HTilde}
\begin{split}
  \tilde{H} \phi(\mathbf{r}) &=\frac{\hbar^2}{2 a^2 m_r} \Big( 6 \phi(\mathbf{r}) \\
  &- \sum_{\substack{i=x,y,z \\ \delta=\pm 1}} \Big[ 1+\frac{2 m_r}{M} (\cos a K_i -1) \Big]^{1/2} \,  \phi(\mathbf{r}+ \delta \mathbf{e}_i) \Big) \\& +  U(\mathbf{r}) \phi(\mathbf{r}) \;.
\end{split}
\end{equation}

For small $ |\mathbf{K}| \ll \pi/a$, we can expand the cosine and square root in $\tilde{H}$ to second order, to find the Hamilton operator for low momentum states
\begin{equation}\label{Hll}
\tilde{H}_{\ll} = 
 \frac{\hbar^2 |\mathbf{K}|^2}{2 M}   + \frac{\hbar^2}{2 a^2 m_r} \sum_{i=x,y,z} \Big[ 1-\frac{m_r a^2 }{2 M} K_i^2 \Big] T_i  + U(\mathbf{r})
\end{equation}
where, similar to Eqs.~\eqref{Te},~\eqref{Th},
the kinetic energy operators 
$T_i \phi(\mathbf{r}) = 2\phi(\mathbf{r}) - \phi(\mathbf{r}+\mathbf{e}_i) - \phi(\mathbf{r}-\mathbf{e}_i)$
are used.

For $\mathbf{K}=0$, the total mass $M$ drops out of the Hamilton operator $\tilde{H}_{\ll}$.
In this respect 
the lattice problem resembles the continuum problem, in so far as only the reduced mass $m_r$ determines the $\mathbf{K}=0$ eigenstates and energies.
Differences may however arise at finite $\mathbf{K}$.

We can now perform the continuum limit for $\tilde{H}_{\ll}$, by letting the effective lattice constant $a \to 0$ for fixed values of the other parameters.
In this limit, the wave function goes over into a continuous function
$\phi_c(a \mathbf{r}) =\phi(\mathbf{r})$.
The kinetic energy operator reduces to the the derivative operator 
$T_i/a^2 \to  - \partial^2/\partial r_i^2$,
and $V$ drops out.
We obtain the continuum Hamilton operator
\begin{equation}\label{Hc}
\tilde{H}_{c} =  \frac{\hbar^2}{2M} |\mathbf{K}|^2 - \frac{\hbar^2}{2 m_r} \nabla^2 - \frac{e^2}{\epsilon r} \;,
\end{equation}
 the Hamilton operator of a hydrogen-like atom.
The eigenenergies for $\mathbf{K}=0$ are $E_n = - R_X/n^2$ with the excitonic Rydberg $R_X$,
and the lowest state wave function radius is the excitonic Bohr radius $a_B$.
Both quantities are given by
\begin{equation}\label{RXAB}
R_X = \frac{m_r e^4}{2 \hbar^2 \epsilon^2} \;, \quad
 a_B=\frac{\hbar^2 \epsilon}{e^2 m_r} \;.
\end{equation}

We choose $R_X$ as the unit of energy and $a_B$ as the unit of length for the lattice problem.
The exciton wave functions and energies $E/R_X$ at $\mathbf{K}=0$ are the eigenfunctions and eigenenergies of the dimensionless Hamilton operator 
\begin{equation}\label{HDimLess}
H_X = \Big(\frac{a_B}{a} \Big)^2 \sum_{i=x,y,z} T_i - 2 \Big(\frac{a_B}{a}\Big) u(\mathbf{r}) \;, 
\end{equation}
where
\begin{equation}
u(0)=V \;, \quad u(\mathbf{r}) = 1/|r| \text{ for } \mathbf{r} \ne 0  \;.
\end{equation}
Only two dimensionless parameters, $a_B/a$ and $V$, occur.
The parameter $a_B/a$ distinguishes between
the lattice regime $a_B/a \lesssim 1$ with significant central-cell corrections
and the continuum -- or Mott-Wannier -- regime $a_B/a \gg 1$, where central-cell corrections are absent.
In the limit $a_B/a \to \infty$ the eigenenergies $E_n$ of this Hamilton operator 
approach the normalized hydrogen values $E_n = -1/n^2$,
independent of $V$.

\subsection{Central-cell corrections and the exciton mass}

The exciton mass is defined as
\begin{equation}
 M_{X}^{-1} = \frac{\partial^2}{\partial K_i^2} E(\mathbf{K}) |_{\mathbf{K}=0} \;,
\end{equation}
where $E(\mathbf{K})$ denotes the exciton binding energy as a function of $\mathbf{K}$ (because of isotropy of the Hamilton operator the result is independent of $i=x,y,z$).
From Eq.~\eqref{Hll} it follows that 
\begin{equation}\label{MExEnh}
 M_{X} = \frac{M}{1- \frac{1}{2} \langle \phi_0, T_i \phi_0 \rangle } \;,
\end{equation}
where $\phi_0(\mathbf{r})$ is the lowest state wave function of $H_X$.
Note that for a (low-energy) bound state $0 < \langle T_i \rangle < 1$.
In general, $M_{X}/M > 1$, and the exciton is heavier than electron and hole combined.
In the continuum limit, $\langle T_i \rangle \to 0$ as the wave function becomes larger, and $M_{X}=M$.

Interestingly, by Eq.~\eqref{MExEnh} the mass enhancement $M_{X}/M$ depends only on the (normalized) kinetic energy in the lowest exciton state at $\mathbf{K}=0$.
While this statement is trivially true in the continuum limit, it is somewhat surprising for the lattice case.
As discussed above, the $\mathbf{K}=0$ wave function depends on $a_B/a$ and $V$ only.
We see again that out of the five basic parameters of the model 
only two combinations determine the importance of central-cell corrections.

Based on the simple cosine dispersion, Eq.~\eqref{MExEnh} is too crude to give accurate results for actual materials, but it can provide us with reasonable estimates.
Consider for simplicity a 3D wave function given as a product of exponentials
$\phi(x) \propto \exp( - a |x|/a_B)$ along each axis (the estimates are independent of dimension).
Then, $\langle T_x \rangle = 2-2/\cosh(a/a_B)$,
or $M_{X}/M = \cosh (a/a_B)$.
For $a_B=a$, we have $M_{X}/M \approx 1.5$,
but already for $a_B=4a$ it is $M_{X}/M \approx 1.03$ close to one.
Basically we see that central-cell corrections on the exciton mass are important for rather small, yet mobile, excitons. We expect that their properties depend strongly on the local Coulomb interaction $V$ in Eq.~\eqref{UCoul}.
Furthermore, the stronger the binding the larger the mass enhancement.

Eq.~\eqref{MExEnh} holds for any form of the interaction potential $U(\mathbf{r})$,
but it fails in the most general case of different electron and hole dispersion.
It has been derived as early as in Ref.~\onlinecite{MG84} --
where it was noted to explain the immobility of Frenkel excitons with high binding energy --, 
but we feel that the explanation given here is simpler.

\section{Method} \label{sec:Method}

In a straightforward computational approach 
numerical eigenfunctions of a lattice Hamilton operator such as in Eq.~\eqref{HTilde} 
are restricted to a cubic box $[1,L]^3$ of finite extension.
This requires storage of wave function values at $L^3$ lattice sites.
The Hamilton operator is given by a sparse matrix, which allows for fast computations with iterative diagonalization~\cite{So02} or spectral algorithms~\cite{WWAF06}.

As discussed in the introduction, the problem with such plain lattice calculations is the growth of the numerical Hilbert space with $L^3$, which prevents calculations for weakly bound or excited states with large radius.
Approaching the continuum limit the effort diverges although the wave function converges to a well-behaved continuous function.
On the other hand, calculations in the continuum limit require some wave function representation at the beginning, since the natural space discretization through the lattice is missing.

\begin{figure}
\begin{center}
\includegraphics[width=0.9\linewidth]{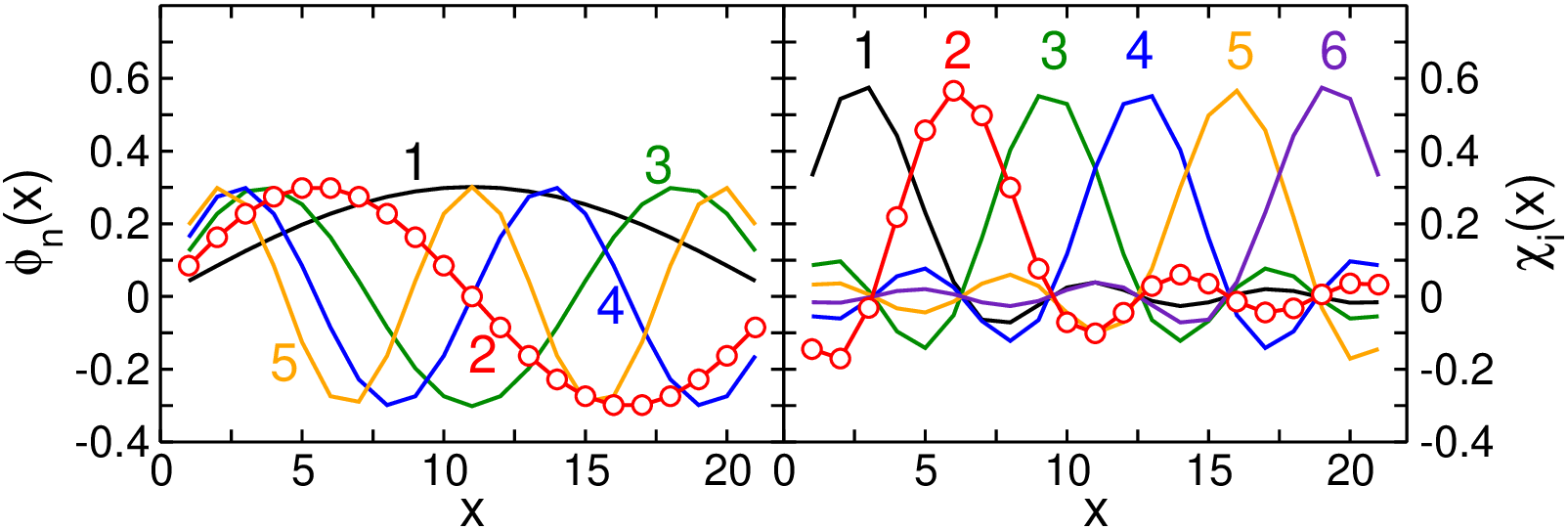}
\end{center}
\caption{First few basis and cardinal functions for the sine basis Eq.~\eqref{Phi} for $L=21$.
Left panel: Basis functions $\Phi_1(x), \dots, \Phi_5(x)$ for $L=21$.
The positions of the lattice sites are indicated through the circles for the $\Phi_2(x)$-curve.
Right panel:
Cardinal functions $\chi_1(x), \dots, \chi_6(x)$ for $M=6$ sampling points.}
\label{fig:PhiChi}
\end{figure}

We address these issues with a variational basis of sine functions.
On the lattice, the 1D basis functions are 
\begin{equation}\label{Phi}
 \Phi_n(x) = \begin{cases}  
 \sqrt{\frac{2}{L+1}} \sin(\frac{ \pi n x}{L+1}) & \text{ if } 1 \le x \le L \;, \\ 
  0 & \text{ otherwise } \;.
   \end{cases} 
\end{equation}
The index $n$ runs from $1$ to the maximal value $L$. 
See Fig.~\ref{fig:PhiChi} for a graphical presentation.
From the 1D functions, 3D basis functions are obtained as tensor products $\Phi_{lmn}(\mathbf{r})=\Phi_l(x) \Phi_m(y) \Phi_n(z)$.
Note that for notational convenience, we count $x$ from $1$ to $L$.
In the numerical calculations, where the potential is strongest at the origin,
we work  on the cube $[-L,L]^3$.

The $\Phi_n(x)$ are orthonormal, with scalar product
\begin{equation}\label{PhiOrtho}
 \langle \Phi_m, \Phi_n \rangle = \sum_{x=1}^{L} \Phi_m(x) \Phi_n(x) = \delta_{mn}  \;,
\end{equation}
but they are not complete since they vanish outside of the interval $[1,L]$.
Bound state wave functions, which decay for large $|x|$ typically exponentially, 
can be represented accurately for sufficiently large $L$.
In our variational calculations we allow the parameter $L$ to grow for minimization of the approximation error.

The $\Phi_n(x)$ have well-defined parity 
\begin{equation}
 \Phi_n(L+1-x) = (-1)^{n+1} \Phi_n(x) \;.
\end{equation}
The matrix elements of the kinetic energy operator are readily calculated as
\begin{equation}\label{Tmat}
 \langle \Phi_m, \hat{T} \Phi_n \rangle = 
 \Big(2- 2 \cos \frac{\pi n}{L+1} \Big) \delta_{mn} \;.
\end{equation}
We see that the sine function are the eigenfunctions of the lattice `particle in a box' problem,
although this is of no further relevance here.
Note that the scaling of the kinetic energy is $1/L^2$, instead of $1/L$ for other functions in a box, as a consequence of the Dirichlet-like boundary conditions 
$\Phi_n(0)=\Phi_n(L+1)=0$.

The continuum version of the sine basis is obtained from the lattice construction
in the limit $L \to \infty$, if the coordinate $x$ is scaled accordingly.
We fix an interval $[0,L_c]$, and 
write the basis functions as $\Phi^c_n(x) = \sqrt{L/L_c} \Phi(x L/L_c)$.
For $L \to \infty$, the basis functions are given by the sine functions
\begin{equation}\label{PhiCont}
  \Phi_n(x) = \begin{cases}  
 \sqrt{\frac{2}{L_c}} \sin(\frac{ \pi n x}{L_c}) & \text{ if } 0 \le x \le L_c \;, \\ 
  0 & \text{ otherwise } \;,
   \end{cases} 
 \end{equation}
where $x$ now is a continuous variable $\in \mathbb{R}$.
The orthonormality condition reads $\int_0^{L_c} \Phi_m(x) \Phi_n(x) = \delta_{mn}$,
and the kinetic energy follows as 
$-\partial_{xx} \Phi_n (x) = \Big( \frac{\pi n}{L_c} \Big)^2 \Phi_n(x)$.

We favor the sine basis over alternative choices because of the simplicity of all relevant expressions, the possibility of using fast Fourier transforms,
and their equivalence to the established sine-DVR~\cite{Ba95} in the continuum limit.

\subsection{Discrete variable representation}

In a variational calculation, the 1D wave function
is given as a linear combination 
\begin{equation}\label{phi}
\phi(x) = \sum_{n=1}^N \phi_n \Phi_n(x)
\end{equation}
of a finite number $N$ of the basis functions (or $N^3$ in 3D).
Note that always $N \le L$.
In the limit $N=L$ the variational basis is complete in the box,
and the use of the sine basis is equivalent to a plain lattice calculation.
Computational savings are expected for $N \ll L$.

While the kinetic energy is diagonal in the sine function basis,
evaluation of the potential term 
requires
multiplication of the coefficients $\{\phi_1, \dots, \phi_n\}$ with a dense matrix $V_{nm}=\langle \Phi_n(x), V(x) \Phi_m(x) \rangle$,
\begin{equation}\label{VDense}
 \tilde{\phi}_n = \sum_{m=1}^N V_{nm} \phi_m \;.
\end{equation}
Formally, this equation defines the projection $\tilde{V}$ of $V(x)$ on the variational Hilbert space spanned by the functions $\Phi_1(x),\dots,\Phi_N(x)$,
with 
\begin{equation}
\tilde{V} \phi(x) = \tilde{\phi}(x) = \sum_{n=1}^N \tilde{\phi}_n \Phi_n(x) \;.
\end{equation}
Note that $\tilde{V}$, in contrast to $V(x)$, is not a multiplication operator in the $x$-eigenbasis. 

In Eq.~\eqref{VDense}, there are $N^2$ matrix elements in 1D, but in 3D the effort grows as $N^6$.
The principal idea of the DVR
to prevent this rapid growth 
is the approximate evaluation of the potential term at the sampling points of a Gaussian quadrature rule~\cite{DC68}.
This is possible if the basis functions are, essentially, polynomials in the position operator $x$.
For the sine basis, we have
\begin{equation}\label{PhiFromP}
  \Phi_n(x) = P_{n-1}(\hat{x}) \Phi_1(x) \;,
\end{equation}
with the transformed lattice position
\begin{equation}\label{hatx}
 \hat{x} = \cos  \frac{\pi x}{L+1} \;,  
\end{equation}
and the Chebyshev polynomials of second kind
\begin{equation}\label{PAcos}
  P_n(x) =  \frac{\sin [(n+1) \arccos x]}{\sin [\arccos x]}  \quad (|x|<1) \;.
\end{equation}
Recall that these polynomials satisfy a three-term recurrence
\begin{equation}\label{PRec}
\begin{split}
  P_0(x)&=1 \;, P_1(x)=2x \;, \\
  P_{n+1}(x)&=2x P_n(x)-P_{n-1}(x) \;.
  \end{split}
\end{equation}
The orthonormality of the basis functions Eq.~\eqref{PhiOrtho} provides the discrete orthogonality relation
\begin{equation}\label{PDiscOverL}
\begin{split}
\delta_{mn} &= \langle \Phi_m(x), \Phi_n(x) \rangle 
 = \langle P_m(\hat{x}) \Phi_1, P_n(\hat{x}) \Phi_1 \rangle \\
& =\frac{2}{L+1} \sum_{x=1}^L \sin^2 \frac{\pi x}{L+1} P_m(\cos  \frac{\pi x}{L+1}) P_n(\cos  \frac{\pi x}{L+1})  
 \end{split}
\end{equation}
for the Chebyshev polynomials.

The sampling points of Gaussian quadrature for the Chebyshev polynomials are the $M$ roots 
\begin{equation}\label{Proot}
 \hat{x}_k = \cos \frac{\pi k}{M+1}   \;,   \qquad 1 \le k \le M \;,
\end{equation}
of the polynomial $P_M(\hat{x})$.
The roots are distinct, and $-1 < \hat{x}_k < 1$.
Note the slight technical complication that $\hat{x}_k$ is given as a transformed position according to Eqs.~\eqref{PhiFromP},~\eqref{hatx}. In original lattice coordinates we have $x_k = k (L+1)/(M+1)$.

Since the variational wave function $\phi(x)$ in Eq.~\eqref{phi} is a linear combination of the first $N$ basis functions, its construction through Eq.~\eqref{PhiFromP}
involves polynomials of maximal degree $N-1$.
The wave function is thus uniquely specified through the values
\begin{equation}
\xi_k = \sum_{n=1}^N \phi_n P_{n-1}(\hat{x}_k)
\end{equation}
at $M$ sampling points $\hat{x}_k$ for every $M \ge N$. 
Note that in general $x_k \notin \mathbb{Z}$, such that $\phi(x_k)$ itself is not defined.

The DVR assumes that instead of multiplication with the dense matrix $V_{mn}$ as in Eq.~\eqref{VDense} the potential term can also be evaluated through the simpler multiplication of the 
wave function values $\xi_k$ with the potential values $V(\xi_k)$ at $M=N$ sampling points,
i.e.
\begin{equation}\label{DVR}
  \tilde{\xi}_k = \sum_{n=1}^N \tilde{\phi}_n P_{n-1}(\hat{x}_k) \approx V(\hat{x}_k) \xi_k \;.
\end{equation}
Recall that $\tilde{\phi}_n$ are the coefficients of the wave function
$\tilde{V} \phi(x)$ in the sine basis from Eq.~\eqref{VDense}.

Eq.~\eqref{DVR} is exact in the full Hilbert space (i.e. for $M=N=L$),
where it just states that the potential acts as a multiplication operator in the position eigenbasis,
i.e. $(V \phi)(x) = V(x) \phi(x)$.
In general, it is an approximation because the projection $\tilde{V}$ of the potential operator onto the variational Hilbert space is no longer a multiplication operator.
Since Gaussian quadrature with $N$ points is exact for polynomials of maximal degree $2N-1$, the approximation is expected to become accurate for sufficiently large $M$, $N$.

The benefits of the approximation Eq.~\eqref{DVR} are two-fold:
First, the potential is now given by a diagonal operation, with $N^3$ instead of $N^6$ effort.
Second, instead of the matrix elements $V_{nm}$ only the function values $V(x_k)$ are needed.
The major drawback is that the approximation violates the variational principle.

\subsection{Variational discrete variable representation}

The DVR uses as many sampling points as basis functions, i.e. $M=N$.
It has been noted~\cite{Fr86,CT95} that the accuracy of the approximation Eq.~\eqref{DVR} can be improved by using larger $M>N$. 
The question then is how large $M$ has to be chosen, e.g. for a singular potential where Gaussian quadrature encounters difficulties.
A nice result, which seems to have been missed in the literature, provides a complete answer: 
Independent of the potential, $M=2N-1$ suffices for the \emph{exact} evaluation of the potential term in the DVR fashion of Eq.~\eqref{DVR}.

The crucial observation is that the projected potential $\tilde{V}$ 
acts on the wave function in the same way as a polynomial in $\hat{x}$ 
\begin{equation}\label{VP}
  V_P(\hat{x}) = \sum_{n=1}^{2N-1} V_{n1} P_{n-1}(\hat{x})  
    \end{equation}
of maximal degree $2N-2$,
with coefficients given by
\begin{equation}\label{Vn}
\begin{split}
  V_{n1} &= \langle \Phi_n(x), \Phi_1(x) V(x) \rangle \\
  &= \frac{2}{L+1} \langle \sin \frac{\pi n x}{L+1} \sin \frac{\pi x}{L+1}  , V \rangle \;.
  \end{split}
\end{equation}
This follows from comparison of matrix elements in the subspace basis $\Phi_1(x), \dots, \Phi_N(x)$.
We first note that 
$V_{mn}= \langle \Phi_m(x), V(x) \Phi_n(x) \rangle
= \langle P_m(\hat{x}) P_n(\hat{x}) \Phi_1^2(x) ,  V(x) \rangle$.
The product $P_m(\hat{x}) P_n(\hat{x})$ is a polynomial of maximal degree $2N-2$,
which is a linear combination of $P_0(\hat{x}), \dots, P_{2N-2}(\hat{x})$.
It thus suffices to compare the $2N-1$ matrix elements $\langle P_n(\hat{x}) \Phi_1^2(x), \dots \rangle$ for $0 \le n \le 2N-2$.
By definition of $V_P(\hat{x})$, we have 
$\langle P_n(\hat{x}) \Phi_1^2(x), V_P(\hat{x}) \rangle = V_{n+1,1} = \langle P_n(\hat{x}) \Phi_1^2(x), V(x) \rangle$, using the orthogonality Eq.~\eqref{PDiscOverL} of the $P_n(\hat{x})$ for the first equality.
This concludes the argument.

Since the true projected potential $\tilde{V}$ can be replaced identically by the polynomial
$V_P(\hat{x})$ for calculations with the variational wave functions,
the potential term can be evaluated exactly through Gaussian quadrature.
To a wave function $\phi(x)$ we associate the polynomial
$\phi_P(\hat{x}) = \sum_{n=1}^N \phi_n P_{n-1}(\hat{x})$,
such that $\phi(x) = \phi_P(\hat{x}) \Phi_1(x)$.
Acting with the potential term $V_P(\hat{x})$ now gives another polynomial
$V_P(\hat{x}) \phi_P(\hat{x})$ of maximal degree $3N-3$.
From this, the polynomial $\tilde{\phi}_P(\hat{x})$ of the new wave function must be obtained,
again of maximal degree $N-1$.
In total, we deal with polynomials of maximal degree $4N-4$.
Since a Gaussian quadrature with $M$ sampling points is exact for polynomials of maximal degree $2M-1$,
the choice $M=2N-1$ guarantees that the expression
\begin{equation}\label{VDVRxi}
 \tilde{\xi}_k = V_P(x_k) \xi_k
\end{equation}
is exact.
We can thus evaluate the potential term exactly by using twice as many sampling points as basis functions, and the values $V_P(\hat{x}_k)$ instead of the potential function values $V(x_k)$.
This provides us with the efficiency benefits of the DVR and preserves the variational principle.

\subsection{Implementation}

Let us now explain how we use the variational DVR (VDVR) in practice.
First note that the wave function coefficients $\phi_n$ and the values $\xi_k$ at the sampling points are related through an orthogonal transformation with the matrix
 \begin{equation}\label{Umat}
 \begin{split}
   U_{nk} &= \lambda_k P_{n-1}(\hat{x}_k) =  \sqrt{\frac{2}{M+1}} \sin \frac{\pi n k}{M+1} \;, \\
   & \text{ with }\lambda_k = \sqrt{\frac{2}{M+1}} \sin \frac{\pi k}{M+1} \;.
  \end{split}
 \end{equation}
A direct calculation shows that $U^+ U = U U^+ = 1$.
The origin of this matrix is that it contains the eigenvectors of $\hat{x}$ in the
basis $\Phi_1(x), \dots, \Phi_{M}(x)$ as the columns.
It is
\begin{equation}\label{lamxi}
\lambda_k \xi_k = \sum_{n=1}^N U_{nk} \Phi_n \;.
\end{equation}
Note that for $M>N$ only a rectangular $N \times M$ submatrix of $U_{nk}$ is used.
Note further that only $L$ basis functions exist on the interval $[1,L]$, such that $M = \min \{ 2N-1 , L \}$.

As a side remark, we mention the eigenfunctions 
$\chi_k(x) = \sum_{n=1}^{M} U_{nk} \Phi_{n}(x)$
of $\hat{x}$ in the Hilbert space spanned by $\Phi_1, \dots, \Phi_{M}$,
the cardinal functions.
They can be used to represent $\phi(x) = \sum_{k=1}^M  \lambda_k \xi_k \chi_k(x)$
directly through the values $\xi_k$ at the sampling points.
Typical cardinal functions are depicted in Fig.~\ref{fig:PhiChi}.
For the maximal value $M=L$, $\chi_k(x) = \delta_{xk}$ is the lattice $\delta$-function localized at the single site $x=k$. We here recover the plain lattice calculation.
We will not use cardinal functions explicitly in this work.

We can now proceed as follows for the evaluation of the potential term:
Obtain the values $\xi_k$ from the coefficients $\phi_n$ through transformation with $U_{nk}$
(Eq.~\eqref{lamxi}),
then multiply each $\xi_k$ with $V_P(\hat{x}_k)$ according to Eq.~\eqref{VDVRxi},
and transform back to the new coefficients $\tilde{\phi}_n$.
That is,
\begin{equation}\label{VDVR}
 \tilde{\phi}_n = \sum_{k=1}^{M} U_{nk} \, V_P(\hat{x}_k) \, \sum_{m=1}^N U_{mk} \phi_m \;,
\end{equation}
or more concisely $\tilde{\phi} = U  V_P U^+ \phi$ where the matrix $V_P \equiv (V_P(\hat{x}_k) \delta_{jk})_{jk}$ is diagonal.
Note that, in contrast to the standard DVR,
we consider the $\phi_n$ instead of the $\xi_k$ as the primary objects in the calculation
since it simplifies the formulation and calculation for $M>N$ (cf. Ref.~\onlinecite{CT95}).

The 1D Hamilton operator in the (V)DVR formulation is given by
\begin{equation}\label{HVDVR}
H_\mathrm{VDVR} =  U V_P U^+ + T \;,
\end{equation}
with the diagonal $N\times N$ kinetic energy matrix $T$ from Eq.~\eqref{Tmat},
the diagonal $M \times M$ potential matrix $V_P$,
and the rectangular $N \times M$ transformation matrix $U$.

The extension to 3D is straightforward.
The Hamilton operator retains the form of Eq.~\eqref{HVDVR}.
The kinetic energy is the sum $T_x + T_y + T_z$ and remains diagonal,
and the diagonal potential matrix has now $N^3$ entries.
The transformation matrix is a tensor product $U^{(3)} = U \otimes U \otimes U$,
which acts as $\xi_{ijk} = \sum_{lmn} U_{il} U_{jm} U_{kn} \phi_{lmn}$.

The multiplication with the $U^{(3)}$ matrix is best done sequentially.
Since in the tensor products each matrix applies only along a single axis, every multiplication requires $N^4$ operations.
The matrices $V$ and $T$ are diagonal, and require $N^3$ operations.
The total operation count is thus of the order $N^4$ instead of $N^6$.
A certain overhead for the additional transformations is present.
With a more detailed counting we find that already for $N=10$ the effort is reduced by a factor $3$ in comparison to a non-DVR evaluation of the potential term.
For $N=100$, the reduction is by a factor of $360$:
A calculation that takes ten minutes with the (V)DVR would take two and a half days without!
The numbers become even more favourably if we use a fast Fourier transform (FFT) for the multiplication with $U^{(3)}$.

\subsection{Calculation of $V_P(x_k)$}

The potential enters Eq.~\eqref{VDVRxi}
through the $M=2N-1$ matrix elements
$V_P(\hat{x}_k)$.
They are obtained from the $V_{n1}$ in Eq.~\eqref{Vn} through transformation with $U_{nk}$ as
\begin{equation}\label{VX}
  V_P(\hat{x}_k) = \sum_{n=1}^{2N-1} P_{n-1}(\hat{x}_k) V_{n1} = \frac{1}{\lambda_k} \sum_{n=1}^{2N-1} U_{nk} V_{n1} \;.
\end{equation}
In practical applications the evaluation of the scalar product in Eq.~\eqref{Vn}, which requires summation over $L$ (or $L^3$) lattice sites, is not desirable.
If the potential is given by a smooth function $V(x)$ we can use Gaussian quadrature instead.
With $N_g \ge M$ sampling points, we obtain the approximation
\begin{equation}\label{VXApp1}
 V_P(x_k) \simeq \frac{1}{\lambda_k} \sum_{n=1}^{M} U_{nk}  \sum_{j=1}^{N_g} U_{nj}^{N_g} \lambda_j^{N_g} V \Big( j \frac{L+1}{N_g+1} \Big) \;,
\end{equation}
where $U_{nj}^{N_g}$, $\lambda_j^{N_g}$ are defined as in Eq.~\eqref{Umat} with $N_g$ replacing $M$.
This expression essentially describes the projection from polynomials of maximal degree $N_g$ onto polynomials of maximal degree $M$, all given through their values at certain sampling points.
Note that the argument of $V(x)$ is not necessarily an integer.
Therefore, we must know $V(x)$ for continuous $x$, not only at the lattice sites.
For $N_g=M$, the Gaussian approximation reduces to the DVR-like expression
\begin{equation}\label{VXApp2}
  V_P(\hat{x}_k) \simeq V\Big( k \frac{L+1}{M+1} \Big) \;, \qquad k=1, \dots, M \;.
\end{equation}

The entries of $V_P$ in Eq.~\eqref{HVDVR} can be obtained through any of the expressions Eqs.~\eqref{VX},~\eqref{VXApp1},~\eqref{VXApp2}.
In practice, we use the simplest approximation Eq.~\eqref{VXApp2} for regular potentials, e.g. (an-) harmonic oscillators. For singular potentials, such as the Coulomb potential, we use Eq.~\eqref{VXApp1} with $N_g = 4 \dots 8 N$, avoiding sampling points at the potential singularity.
On the lattice we treat localized ($\delta$-function) contributions, e.g. from the $V$-term in Eq.~\eqref{UCoul}, exactly through Eq.~\eqref{VX}, and use again Eqs.~\eqref{VXApp1},~\eqref{VXApp2} for the remaining long-range part of the potential.

 \subsection{Discussion}

An important conceptual difference between the VDVR and the original DVR formulation is the separation of the calculation of potential matrix elements through Eq.~\eqref{VX} and their actual usage in Eqs.~\eqref{VDVR},~\eqref{HVDVR}.
Two sources of error exist in the DVR:
First, the replacement of the full, dense matrix-vector multiplication Eq.~\eqref{VDense}
through the diagonal expression Eq.~\eqref{DVR}.
This error is completely eliminated in the VDVR Eqs.~\eqref{VDVRxi},~\eqref{VDVR}
through the choice of $M=2N-1$ sampling points.
Second, the error incurred through approximate evaluation of the matrix elements $V_P(\hat{x}_k)$.
This error can be made smaller through better approximations to Eq.~\eqref{VX},
e.g. by increasing $N_g$ in Eq.~\eqref{VXApp1}.

In the original DVR, both errors contribute equally.
The obvious way for error reduction is to increase $N$,
which  inflicts a computational overhead on the entire calculation.
In the VDVR, the effort for a better calculation of the $V_P(\hat{x}_k)$ needs to be invested only once in Eq.~\eqref{VX} or Eq.~\eqref{VXApp1} before the actual use of the Hamilton operator Eq.~\eqref{HVDVR},
but there is no reason for increasing $M$ beyond $2N-1$ for the evaluation of the potential term.

In our opinion this is a central advantage of the formulation chosen here:
The calculation of matrix elements of $V(x)$ is completely independent of the Gaussian quadrature underlying the diagonal VDVR evaluation of the potential term.
In particular for singular potentials a large number $N_g$ of sampling points in a Gaussian quadrature Eq.~\eqref{VXApp1} can be necessary to obtain accurate matrix elements,
while the variational wave function is a good approximation already for $N,M \ll N_g$.
It can also be useful to calculate the $V_P(x_k)$
with other integration/summation procedures,
e.g. adaptive Gauss-Kronrod-integration or specialized routines for functions with an integrable singularity.
In the original DVR, replacing the $V(x_k)$ in Eq.~\eqref{DVR} by better matrix elements is not possible, and increasing $N$ is the only possibility for improvement.

\subsection{Examples}\label{sec:Cont}

 \begin{figure}
\begin{center}
\includegraphics[width=0.9\linewidth]{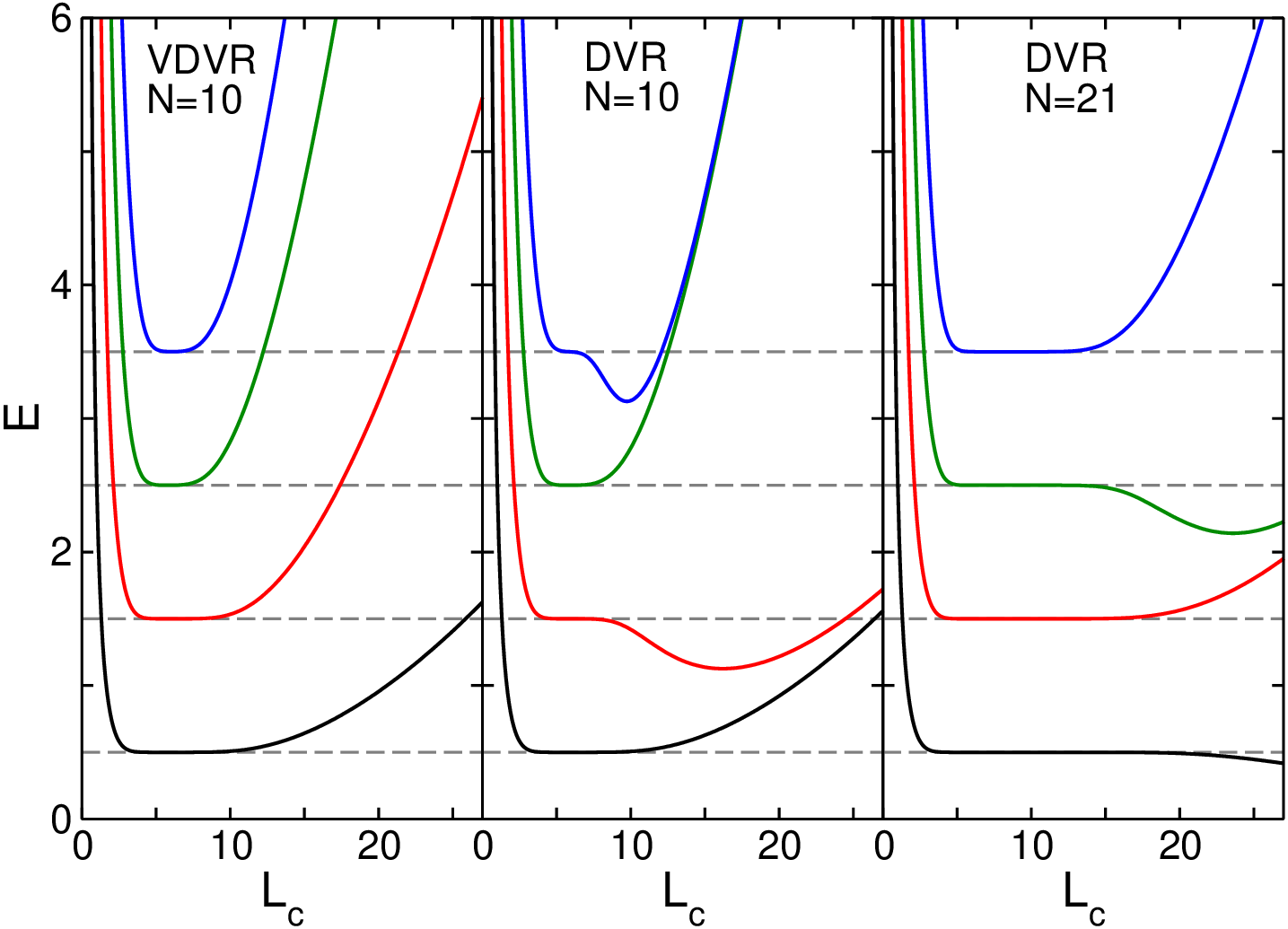}
\end{center}
\caption{Numerical energies of the four lowest eigenstates of the harmonic oscillator $H = -\partial_{xx}/2 + x^2/2$, as a function of the interval length $L_c$ for the interval $[-L_c,L_c]$.
Shown are results for the VDVR (left panel) and the DVR (central and right panel),
for $N=10, 21$ basis functions as indicated. The exact eigenenergies $E_n = n-1/2$ are indicated by dashed horizontal lines (note that we start counting with $n=1$).
The Gaussian approximation Eq.~\eqref{VXApp2} for the potential matrix elements has been used.}
\label{fig:HOsci1}
\end{figure}

\begin{table}
\begin{tabular}{cllll}
$N$ & $E_1$ & $E_5$ & $E_{10}$ & $E_{20}$  \\\hline
   5    &  0.5001685777  & 5.088729909 & --- & --- \\  
   10  &  0.5000004143  & 4.510831820 & 11.14028090 & --- \\
   20  &  0.5000000000  & 4.500000074 & 9.500334454 & 23.96785196 \\
   30  &  0.5000000000  & 4.500000000 & 9.500000005 & 19.51678526 \\
   40  &  0.5000000000  & 4.500000000 & 9.500000000 & 19.50000254 \\
   50  &  0.5000000000  & 4.500000000 & 9.500000000 & 19.50000000
\end{tabular}
\caption{Convergence of the numerical energies with the number of basis states $N$ in a VDVR calculation, for  the eigenenergies $E_1, E_5, E_{10}, E_{20}$ of the harmonic oscillator $H = -\partial_{xx}/2 + x^2/2$ as in Fig.~\ref{fig:HOsci1}.
The Gaussian approximation Eq.~\eqref{VXApp2} for the potential matrix elements has been used. Note that the energy minimum is obtained for different optimal $L$ (not shown), 
in particular for small $N$.
}
\label{tab:HOsci1}
\end{table}

\subsubsection{Harmonic oscillator}
The difference between the variational and non-variational DVR is apparent
for the (continuum) harmonic oscillator $H = -\frac{1}{2} \partial_{xx} + \frac{1}{2}Êx^2$,
with eigenvalues $E_n = n-1/2$ (note that we start counting with $n=1$).
In Fig.~\ref{fig:HOsci1} we show the numerical result for the energy $E_1,Ê\dots, E_4$ of the four lowest eigenstates under variation of the parameter $L_c$.
Note that the Gaussian approximation Eq.~\eqref{VXApp2} for the potential  is used.
The error has two sources: The domain truncation error because the variational wave function vanishes outside an interval of length $L_c$,
and the basis error from the approximate representation of the wave function through a finite number of basis functions.
Since the VDVR is variational by construction, the numerical energies approximate the true energies from above (left panel).
The optimal value is found by minimization of the respective numerical energy under variation of $L_c$.
The values in Table~\ref{tab:HOsci1} demonstrate the attainable precision.
With only $N=30$ the first 10 eigenvalues are converged,
and even the 20th eigenvalues $E_{20}$ is significant with a relative error below $10^{-3}$.

The standard DVR violates the variational principle even in this simple example (central and right panel in Fig.~\ref{fig:HOsci1}), and it is not recovered for large $N$.
Nevertheless the DVR provides meaningful results if the plateau region is identified, where the DVR energy is almost constant under variation of $L_c$ and a good approximation to the true energies.
But the violation of the variational principle complicates the application of the DVR for our purposes, since an automatic identification of the plateau is difficult and error-prone.
Also, we normally have no a-priori estimate for a suitable $L_c$,
since we do not know the wave function in advance.
 
\subsubsection{Hydrogen atom}
The second example concerns already the numerical calculations for the 3D exciton problem.
We consider the dimensionless Hamilton operator Eq.~\eqref{HDimLess}
for $a_B/a \gg 1$, where plain lattice calculations become increasingly demanding.
In Table~\ref{tab:3DHydro} we show the energy of the lowest state in the continuum limit $a_B/a=\infty$, i.e. for a 3D hydrogen atom, obtained with VDVR.
The convergence to the exact value is not as favourable as for the harmonic oscillator (cf. Table~\ref{tab:HOsci1}), as a consequence of the singular potential.
The singularity of the potential increases the error of the Gaussian approximation Eq.~\eqref{VXApp2}.
The convergence improves for more accurate potential matrix elements,
obtained with Eq.~\eqref{VXApp1} for $N_g=8N$.
Since the additional effort for a better calculation of the matrix elements is only invested once,
prior to the actual use of the VDVR Hamilton operator in an iterative diagonalization procedure,
the running time of calculations increases only marginally (about $5\%$).

A second error source is intrinsic to the sine basis construction, which has difficulties to resolve the cusp of the hydrogen wave function at $\mathbf{r}=0$.
Recall that we approach the continuum limit starting from a cubic lattice.
In the continuum limit, rotational symmetry allows for the separation of radial and angular coordinates and the choice of a better basis, e.g. of Laguerre polynomials.
Starting from the lattice, this is prevented by the reduced lattice symmetry.
Convergence of expansions of a rotationally invariant function with a cusp 
in a basis without rotational symmetry is relatively slow.
Since full rational symmetry is restored only in the continuum limit, 
there is no easy fix to this problem.
For smaller $a_B/a$, away from the continuum limit, the error is reduced.

Despite these complications, the error with $N=100$ is smaller than $3 \times 10^{-4}$.
The variational Hilbert space has dimension $10^6$.
As a consequence of parity symmetry, only half of the basis functions along each axis are used, reducing the necessary effort by a factor of $2^3=8$. Making use of the full lattice symmetry requires manipulation of
only $\approx 21000$ states for $N=100$ (effectively, $30$ per coordinate axis).
There is probably room for improvements of the basis, but we are not aware of a simple solution to the lattice/continuum symmetry mismatch problem.
With the present construction, we achieve an error below $10^{-3}$ in all quantities and for all parameters, at modest computational effort.
For $N=30 \dots 50$, the dimension of the variational Hilbert space is $125000$ at maximum without consideration of lattice symmetry. Such calculations, in double real precision, require less than $25$ MB of computer memory. Implementing the full lattice symmetry
reduces these numbers to $2600$ states and about 2 MB of central storage.
Recall that also the running time of a program is significantly reduced  due to the favourable (V)DVR scaling $\propto N^4$.
For the calculations reported here, this level of accuracy and efficiency is sufficient.

\begin{figure}
\begin{center}
\includegraphics[width=0.9\linewidth]{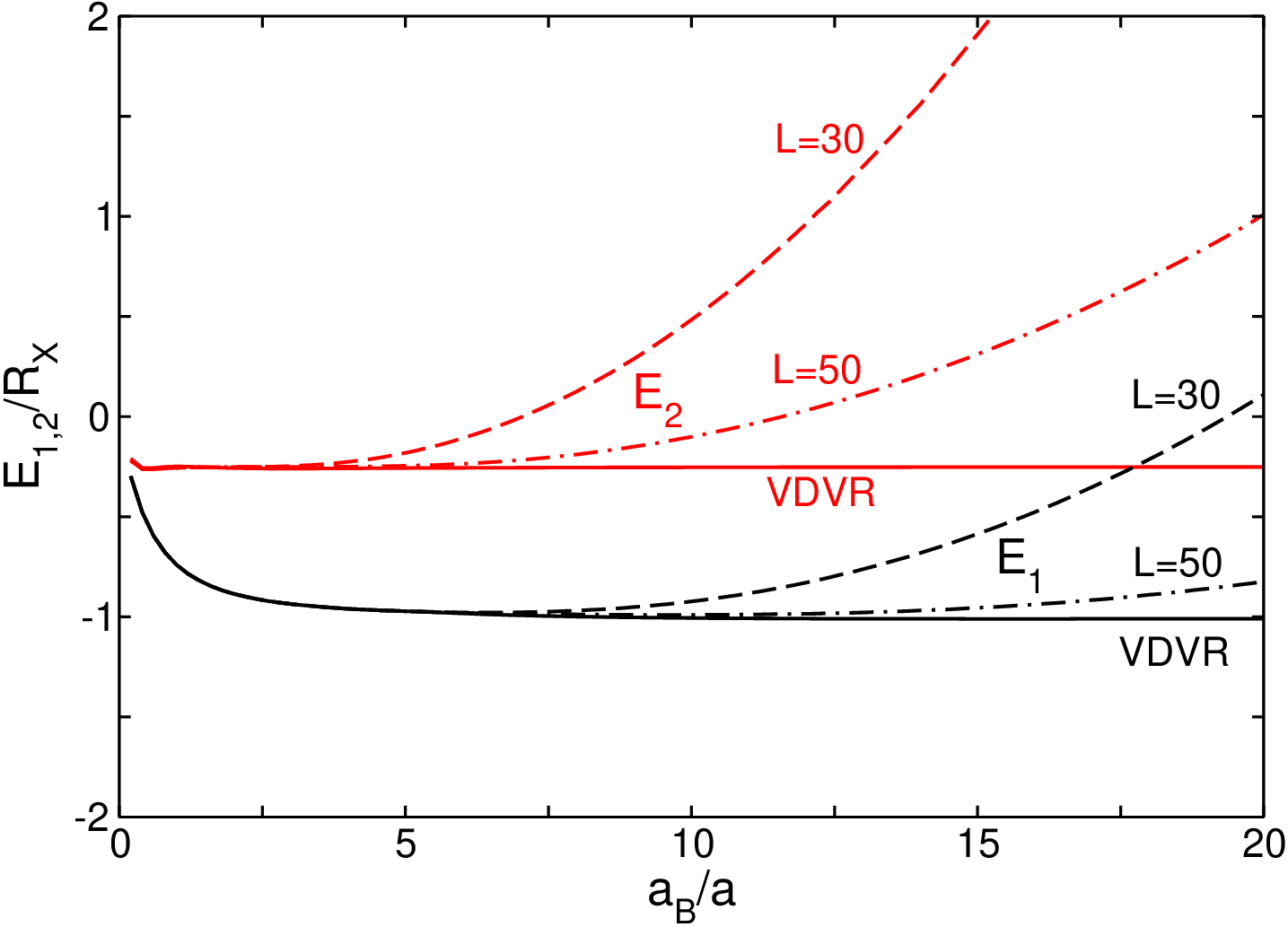}
\end{center}
\caption{Energy of the lowest ($E_1$) and first excited state ($E_2$) for the Hamilton operator Eq.~\eqref{HDimLess} (with $V=1$), as a function of $a_B/a$.
The solid curves are calculated with the VDVR  and $N=30$ basis states.
The dashed curves have been obtained with a plain lattice calculation on a cube $[-L,L]^3$ with $L=30, 50$.}
\label{fig:3DHydro}
\end{figure}

The real benefits of the VDVR over a plain lattice calculation become apparent in Fig.~\eqref{fig:3DHydro}, where we show the energy of the lowest and first excited state of $H_X$ (Eq.~\eqref{HDimLess}) as a function of $a_B/a$.
We compare the VDVR with $N=30$ basis states per direction to a plain lattice calculation on a cube $[-L,L]^3$ with fixed $L=30$.
 For small $a_B/a$, the wave function radius is small and the VDVR reduces to the plain lattice calculation.  As the wave function radius grows with $a_B/a$, the domain truncation error of the plain lattice calculation becomes severe and renders the results meaningless.
 In particular excited states are not accessible.
 On the other hand, the error of the VDVR is bounded independently of the actual extension of the wave function,  and the energies converge to the correct values $E_1/R_X=-1$, $E_2/R_X=-0.25$ for $a_B/a \to \infty$.
 The error in this limit can be deduced from Table~\ref{tab:3DHydro}.

It should be noted that the optimal $L$ in the VDVR is different for the lowest state and excited states, reflecting the larger radius of the latter.
It is the advantage of the variational procedure to adapt itself to these differences.

\begin{table}
\begin{tabular}{cll}
$N$ & $E_1$ & $E_1$ ($N_g=8N$) \\\hline
10 & -0.92931387 & -0.96164993   \\
20 &  -0.97321483 & -0.99026343 \\
30 & -0.98536003 &  -0.99589247  \\
40 &  -0.99059861 & -0.99781906 \\
50 & -0.99338044 &  -0.99871312 \\
60 & -0.99492841&   -0.99912287 \\
70 & -0.99605729 &  -0.99938235 \\
80 & -0.99683500 & -0.99954469 \\
90 & -0.99739575 & -0.99965255 \\
100 & -0.99781451 & -0.99972716 \\
exact & -1.0 \hspace{4em} \\
\end{tabular}
\caption{Convergence of the ground state energy $E_1$  with the number of basis states $N$ in a VDVR calculation, for the Hamilton operator~\eqref{HDimLess}  in the continuum limit $a_B/a = \infty$, i.e. a 3D hydrogen atom.
The values in the 2nd column are obtained with the Gaussian approximation Eq.~\eqref{VXApp2},
the values in the 3rd column from Eq.~\eqref{VXApp1} with $N_g = 8 N$.
}
\label{tab:3DHydro}
\end{table}

\section{Excitons}\label{sec:Exciton}

\begin{figure}
\begin{center}
\includegraphics[width=0.9\linewidth]{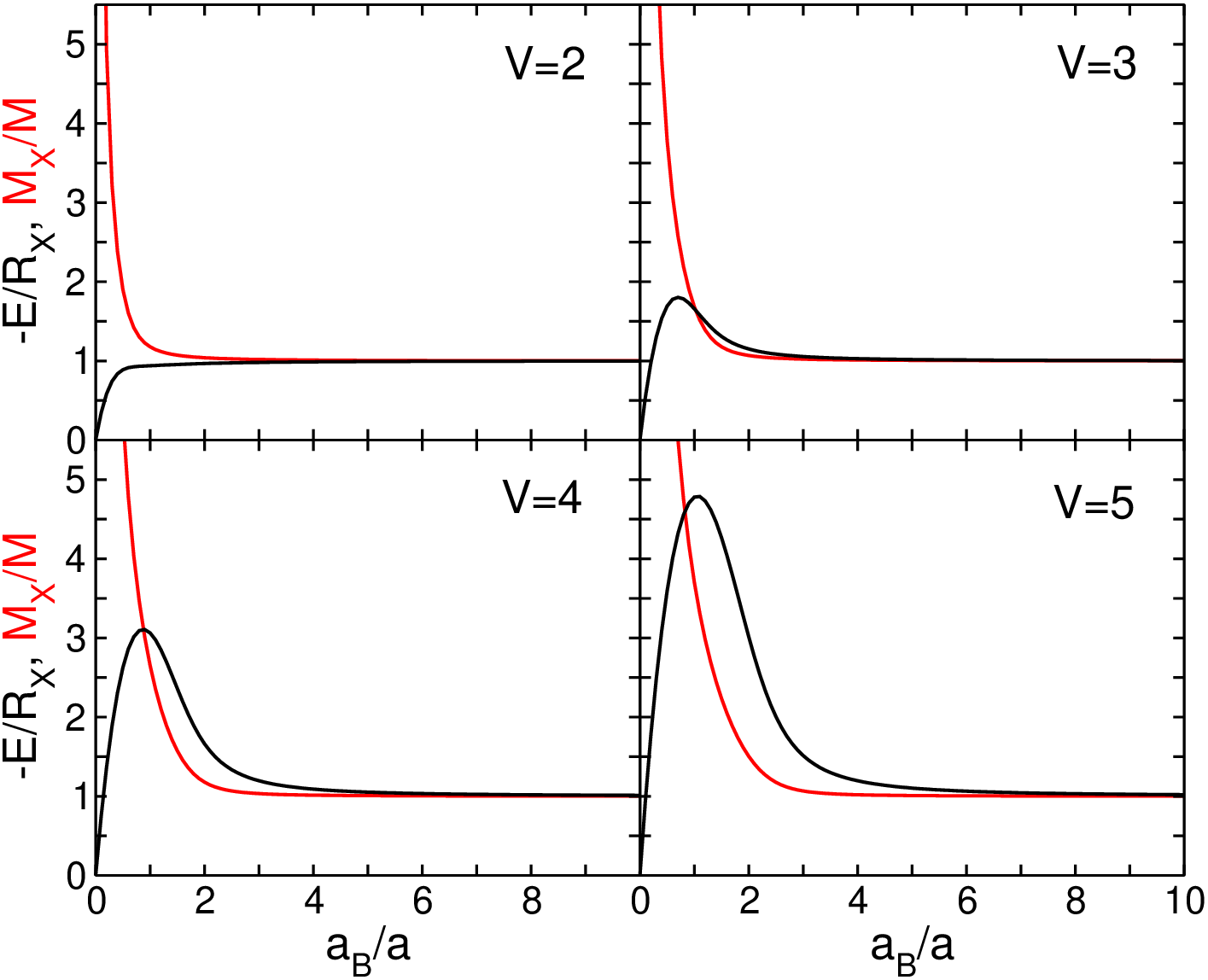}
\end{center}
\caption{Exciton binding energy $E/R_X$ and mass enhancement $M_{X}/M$ as a function of $a_B/a$, for different $V$ as indicated.
Results were obtained with a VDVR calculation for $N=20 \dots 50$
with an error below $10^{-3}$ for both quantities.}
\label{fig:Exc}
\end{figure}

According to Eq.~\eqref{MExEnh} 
and the simple estimate we gave in Sec.~\ref{sec:Model},
central-cell corrections of the exciton mass become important 
if the exciton Bohr radius is of the order of the (effective) lattice constant, 
i.e. for $a_B/a \lesssim 1$.
In Fig.~\ref{fig:Exc} we show the exciton binding energy $-E/R_X$ and the mass enhancement $M_{X}/M$ for the lowest exciton state,
as obtained with the VDVR applied to the Hamilton operator Eq.~\eqref{HDimLess}.
As we discussed in Sec.~\ref{sec:Model}, these curves depend only on the parameters $V$ and $a_B/a$.

As expected, $-E/R_X, M_X/M \to 1$ in the continuum limit $a_B/a \to \infty$, independent of $V$.
Deviations arise for smaller $a_B/a$.
It may be interesting to note that the binding energy changes more pronouncedly with $V$,
while the mass enhancement, as a function of $a_B/a$, remains similar.
The numerical data show that two opposite situations are possible:
A small binding energy and large mass enhancement for small $V$ and $a_B/a$,
and a large binding energy and small mass enhancement for large $V$ and moderate $a_B/a$.
Physically, the magnitude of both parameters $V$, $a_B/a$ is related, e.g., to the extension of Wannier functions for the conduction and valence band.
For reasonable parameters
a large exciton mass coincides with a higher binding energy, and vice versa.

\begin{figure}
\begin{center}
\includegraphics[width=0.9\linewidth]{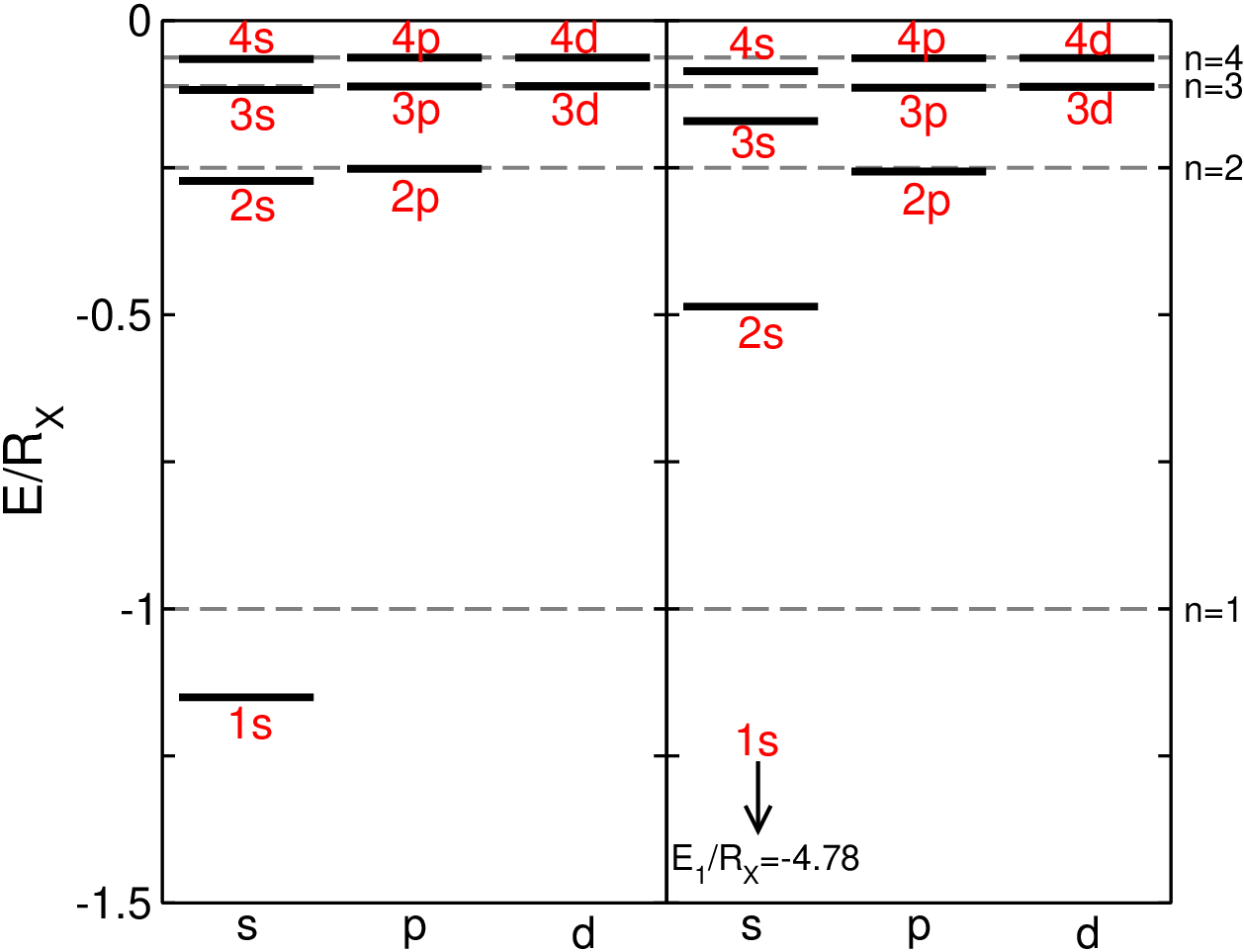}
\end{center}
\caption{Spectrum of the exciton model Eq.~\eqref{HDimLess},
for $V=3$, $a_B/a=2$ (left panel) and $V=5$, $a_B/a=1$ (right panel),
calculated with VDVR.
Grey dashed lines indicate the Rydberg series $E_n/R_X = - 1/n^2$.
}
\label{fig:ExcSpec2}
\end{figure}

\begin{figure}
\begin{center}
\includegraphics[width=0.9\linewidth]{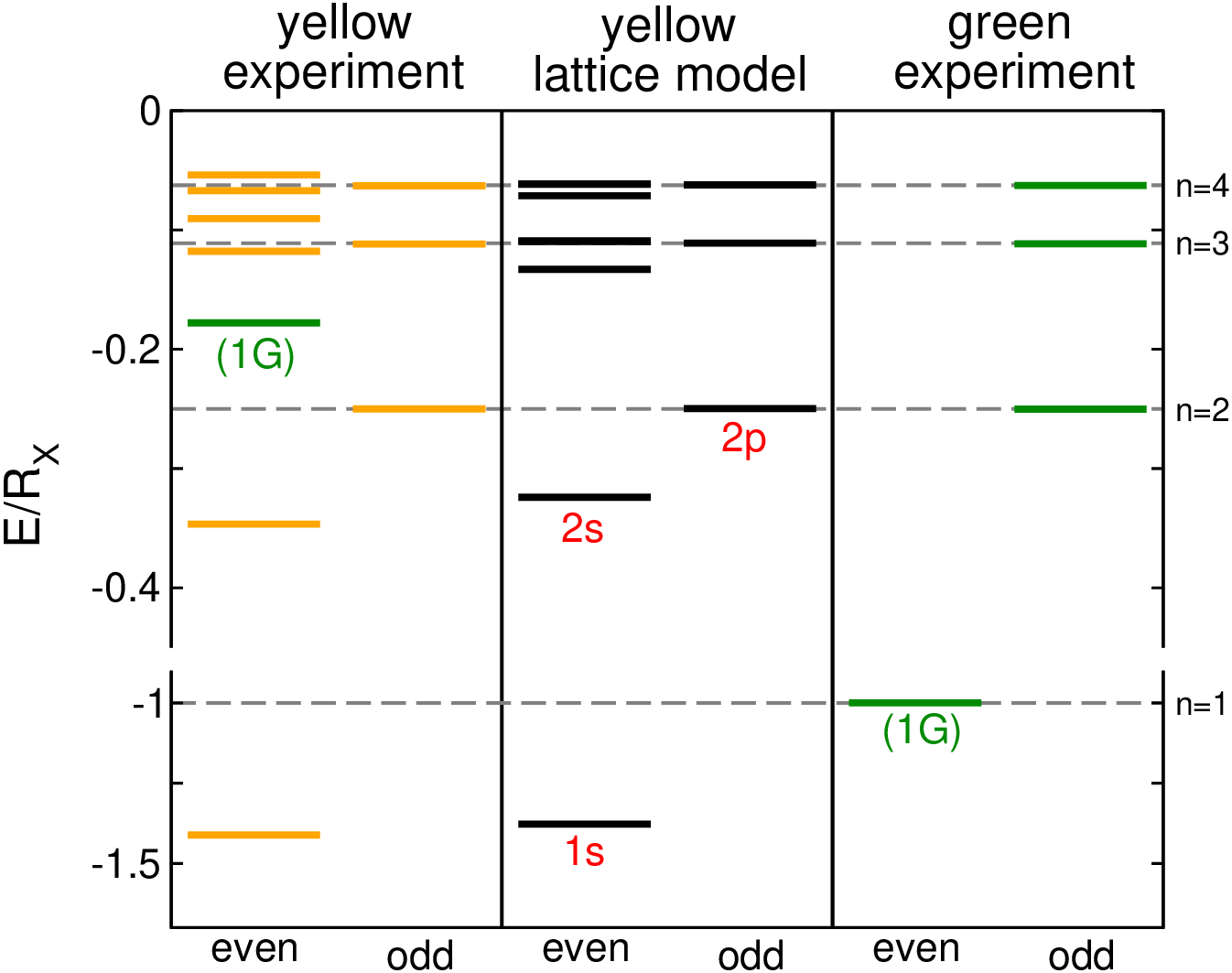}
\end{center}
\caption{Spectrum of the exciton model Eq.~\eqref{HDimLess} in comparison to experimental values for the yellow (ortho-) exciton series in cuprous oxide~\cite{FKUS79,UFK81}.
The model parameters $V=2.75$, $a_B/a=0.98$ are determined from the binding energy and mass enhancement of the lowest (1s) exciton state.
Shown are even and odd parity states for the yellow (left panel) and green (right panel) exciton series, and the results of the VDVR model calculation for the yellow series (central panel).
Note the broken energy axis.
The grey dashed lines give the energies $-1/n^2$ of the Rydberg hydrogen series.
The ``1G'' line in the left/right panel is at identical energy.}
\label{fig:ExcSpec}
\end{figure}

In Fig.~\ref{fig:ExcSpec2} we show the typical change of the exciton spectrum (at $\mathbf{K}=0)$ in comparison to the  
hydrogenic Rydberg series $E_n = -R_X/n^2$, which is realized in the continuum limit $a_B/a \to \infty$.
Starting from there, the $n^2$-fold degeneracy of the hydrogen eigenstates is lifted 
in the lower crystal symmetry.
While the notation in the figure refers to the hydrogen problem,
a group-theoretical classification is possible with the irreducible representation of
the point group $O_h$ of our lattice model, the symmetry group of a cube~\cite{DDJ08}.
The even parity ``s'' states (odd parity ``p'' states) arise from hydrogen states with angular momentum $l=0$ ($l=1$), and correspond to one-dimensional (three-dimensional) irreducible representations.
For the even parity ``d'' states, the $2l+1=5$ dimensional irreducible representation of the full rotation group splits into a two and a three dimensional representation under the reduced symmetry of $O_h$. 
Numerically we see indeed that each such state is a doublet of a two and three-fold degenerate eigenenergy, but the splitting of the order $10^{-4}$ is not resolved in Fig.~\ref{fig:ExcSpec}.

\begin{figure}
\begin{center}
\includegraphics[width=0.9\linewidth]{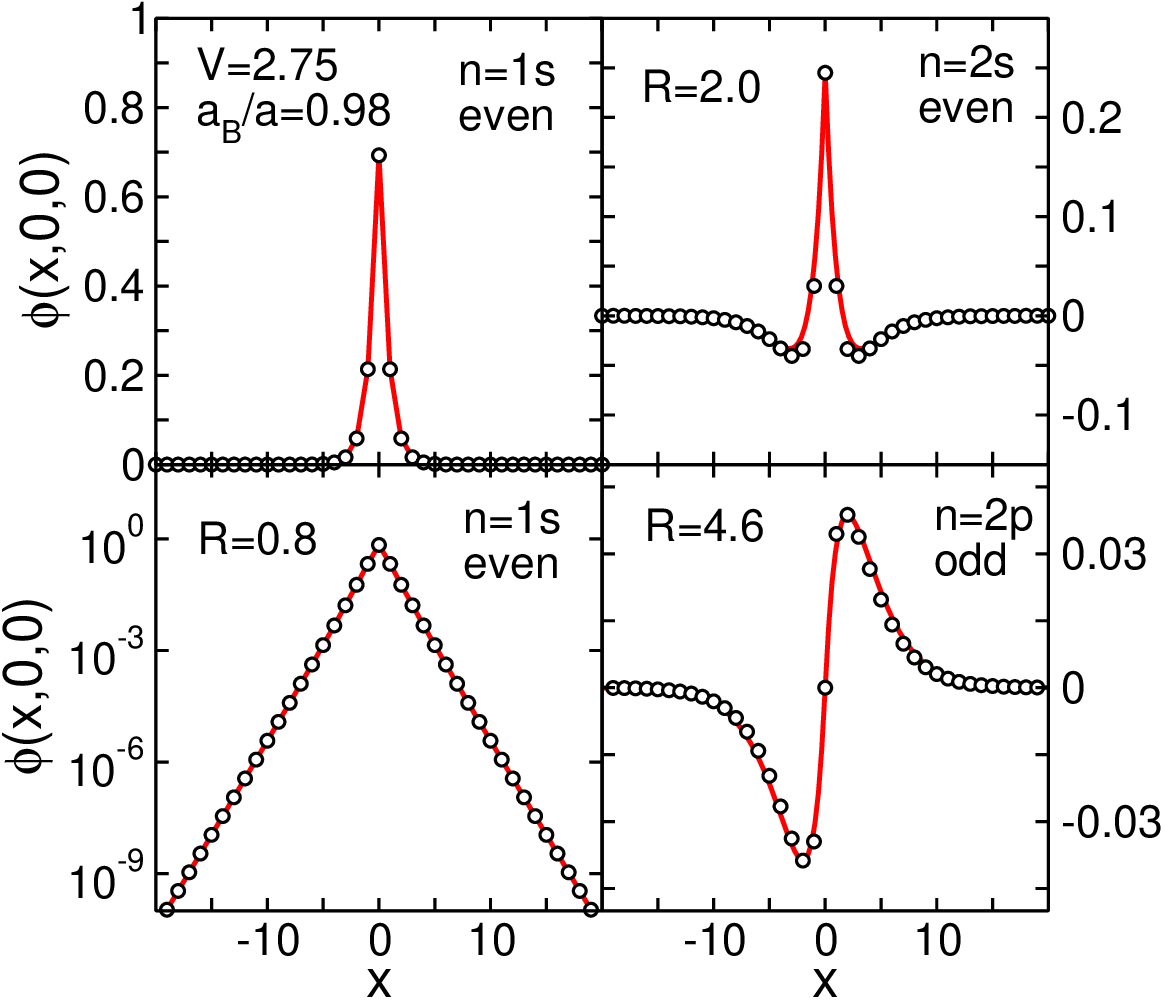}
\end{center}
\caption{Wave function $\phi(x,0,0)$ 
for parameters $V=2.75$, $a_B/a=0.98$ as used for the cuprous oxide in Fig.~\ref{fig:ExcSpec} below.
Shown is the lowest exciton state (left panels, with linear and logarithmic ordinate),
the second even (top right panel) and the first odd state (bottom right panel).
The circles give the values $\phi(x,0,0)$ at the respective lattice site,
the solid red curves show the corresponding hydrogen wave functions for comparison.
Also given is the wave function radius $R$.
}
\label{fig:Wave}
\end{figure}

The wave functions given in Fig.~\ref{fig:Wave} still resemble hydrogen wave functions, although their properties, i.e. the binding energy or mass, do not.
 In our simple model, the energy shifts are induced by the $V$-term in Eq.~\eqref{UCoul},
which affects the ``s'' states strongly since the probability $|\phi(0)|^2$ of electron and hole being in the same unit cell is large.
The ``p,d,\dots'' states are much less affected (for hydrogen states, $\phi(0)=0$ exactly for $l \ge 1$).
Significant energy shifts thus arise from lifting of the dynamical degeneracy of the $1/r$-Coulomb potential with respect to the angular quantum number $l$, 
and are not associated with the splitting of states with different magnetic quantum number $m$ that is predicted by the lower crystal symmetry.

To understand the significance of these results for the cuprous oxide $\mathrm{Cu}_2\mathrm{O}$, we show a model calculation for the yellow (ortho-) exciton series in this material in Fig.~\ref{fig:ExcSpec} in comparison to experimental data from Refs.~\onlinecite{FKUS79,UFK81}.
The experimental absorption energies of the odd parity states can be fitted to a perfect Rydberg series $E^{ye}_n = E^{ye}_g - R^{ye}_X/n^2$, with $E^{ye}_g=2.17 \mathrm{eV}$ for the gap energy and $R^{ye}_X=98.4 \mathrm{meV}$ for the excitonic Rydberg.
With these values, we obtain the left panel in Fig.~\ref{fig:ExcSpec} for the normalized energies
$E/R_X \hat{=} (E^{ye}-E^{ye}_g)/R^{ye}_X$.
Obviously, the energies of the even parity states differ significantly from the Rydberg energies $-1/n^2$.

For the model calculation, 
we choose the two parameters $V=2.75$ and $a_B/a=0.98$ according to a fit of the numerical binding energy and mass enhancement of the lowest (1s) exciton state, as given in Fig.~\ref{fig:Exc}, to the experimental values
$E=139 \mathrm{meV}$ and $M_X/M=1.5$.
Nothing was assumed or adjusted for higher exciton states.
Both parameter values have a reasonable order of magnitude.
Recall that $a_B/a < 1$ is an indication of significant central-cell corrections.

The central panel in Fig.~\ref{fig:ExcSpec} shows the calculated exciton spectrum for the above model parameters.
The spectrum is qualitatively correct, and we find excellent quantitative agreement for the 2s state which reproduces the experimental energy with an error of only $5\%$.
Deviations occur for higher even parity states, which shows the limitations of the simplistic model used here.

The state labelled (1G) in the left panel does not fit into the model spectrum,
and should probably be incorporated into the green exciton series shown in the right panel in Fig.~\ref{fig:ExcSpec}.
Again from a fit of the experimental energies of the odd parity states to a Rydberg series,
resulting in $E^{gr}_g=2.31 \mathrm{eV}$ and $R^{gr}_X=151 \mathrm{meV}$,
the (1G) state is found exactly at the energy $E^{gr}_g - R^{gr}_X$ of the lowest (1s) green exciton state. There is further experimental evidence about the assignment of states to the yellow or green exciton series, such as response to strain and magnetic fields~\cite{TCB81,FKUS79}, but with the present simple model we are unable to provide further analysis.

\section{Conclusions}\label{sec:Conc}

In the present paper 
we introduce a variational discrete variable representation for bound states on a lattice 
and apply it for a study of excitons with significant central-cell corrections.

In the VDVR wave functions are given in a variational basis of sine functions.
It combines
(i) accuracy because of the use of exact matrix elements and the variational determination of the optimal basis function width,
with 
(ii) efficiency,  since it evaluates the potential term in the DVR spirit through a diagonal matrix.
We adapted the original DVRs in two aspects:
Our construction 
(iii) fully preserves the variational principle, and (iv) bridges the gap between lattice and continuum calculations in a single unified framework.

The example of central-cell corrections for excitons 
provides the physical motivation for the present work.
From the simple two-band lattice model adopted here
we can mainly draw qualitative conclusions.
In the regime $a_B \simeq a$ where central-cell corrections become important, excitons are still closer to Mott-Wannier excitons than to Frenkel excitons.
Their properties, however, deviate significantly from the hydrogen picture -- even for excited states where the radius exceeds the lattice constant. 

Despite the simplicity of the two-band model, we can successfully reproduce the experimental spectrum of the yellow exciton series in the cuprous oxide $\mathrm{Cu_2O}$,
even with quantitative agreement.
Only two parameters enter the model calculation, which are fixed by the binding energy and mass of the yellow 1s exciton state.
That the spectrum of excited states can be reproduced from two elementary properties of the lowest exciton state -- one being the exciton mass with no apparent relation to energies of optical transitions -- shows how the use of a lattice model allows us to connect different properties through a fundamental microscopic description. 
The additional experimental information from measurements of the exciton mass can thus be used in a theoretical interpretation of the exciton spectrum.
%%%

The present model calculation is too simplistic to cover all relevant aspects of exciton formation. 
The study of many important effects, such as the spin-dependent energy splitting between ortho- and para-excitons or the influence of electron-hole exchange interaction on the exciton mass~\cite{YIKG07}, has to be postponed to a forthcoming publication, where we will also discuss how the large central-cell corrections for yellow excitons in comparison to the negligible corrections for the green exciton series are related to the respective valence band dispersion.

We will consider extensions of the present work in three directions.
First, refinements of the two-band model are necessary.
As a few principal issues, we can list
(a) the inclusion of realistic band structures, e.g. from ab-initio calculations~\cite{FSSR09},
(b) full consideration of lattice symmetries, which is particularly important for the classification of excited states~\cite{DDJ08},
(c) improved matrix elements for the short-range Coulomb interaction,
which can be principally obtained from the Wannier functions of conduction and valence bands,
and (d) the corrections to the dielectric constant for screening at short distances~\cite{Pe62,Re77}.
The exciton spin configuration is relevant in
(e) the exchange interaction, leading to a splitting of exciton states~\cite{DFKSBS04},
and (f) spin-orbit coupling.
For comparison with experiment it is desirable to allow for
(g) external (magnetic) fields and (h) strain/lattice deformation~\cite{WPBC80,TCB81}.
Also such a refined exciton model can be studied within the VDVR.

More demanding would be the inclusion of dynamical screening,
which is possible within a Green function formalism.
We have discussed elsewhere the use of polynomial bases for Green function calculations~\cite{AF08}, but the combination with VDVR is not worked out.
It would give a polynomial basis construction both for position and energy.
Note that the present work can be understood as the solution of the Bethe-Salpeter equation in the special case of a non-frequency dependent interaction.
We see no urgency to include dynamical screening into the model, because its effect is less significant than the corrections listed above.

Second, our derivation of the VDVR generalizes to arbitrary basis sets of orthogonal polynomials with only minor modifications.
In particular in the continuum limit, where we have more freedom for the basis choice,
such a generalization display its full strength in comparison to artificial ad-hoc discretizations of position or momentum space. 
It will be discussed elsewhere.

Third, the VDVR is powerful enough to allow for the study of biexcitonic systems and exciton-exciton scattering.
Our discussion of the exciton spectrum shows why lattice models are important for an understanding of small-radius excitons, and the same is true for interacting two-exciton systems.
A calculation of the central-cell corrections for exciton scattering lengths is of immediate relevance for Bose-Einstein-condensation.

Bearing in mind that the cuprous oxide is one material where the search for Bose-Einstein-condensation of excitons is justified, microscopic studies of excitons in this material 
always provide us, apart from our genuine interest in excitons away from the Mott-Wannier limit,
with a perspective on the conditions and limitations of collective exciton behaviour.
In this sense, the VDVR technique and the lattice calculations reported here are one building block for the understanding of recent and future experiments on the cuprous oxide and similar materials.

\acknowledgments
We appreciate helpful discussions with H. Stolz on excitons in the cuprous oxide.
This work was supported by Deutsche Forschungsgemeinschaft through \mbox{AL1317/1-1} and
SFB652.

\end{document}